\documentclass[preprint]{revtex4-2}

\usepackage{graphicx}
\usepackage{dcolumn}
\usepackage{bm}
\usepackage{amsmath,amssymb}
\usepackage{braket}
\usepackage[hidelinks]{hyperref}
\usepackage{tabularx}
\usepackage{placeins}
\usepackage{dblfloatfix}
\usepackage{array}
\usepackage{enumitem}
\usepackage{ragged2e}

\newcolumntype{Y}{>{\raggedright\arraybackslash}X}
\newcolumntype{P}[1]{>{\raggedright\arraybackslash}p{#1}}

\begin{document}
	
	\title{Beyond the HOM Dip: A Multi-Metric Module for Teaching Two-Mode Quantum Optical Interference}
	
	\author{Melissa Coronado-Arrieta}
	\affiliation{Department of Physics, New Mexico State University, Las Cruces, New Mexico 88003, USA}
	
	\author{Boris Kiefer}
	\email{bkiefer@nmsu.edu}
	\affiliation{Department of Physics, New Mexico State University, Las Cruces, New Mexico 88003, USA}
	
\begin{abstract}
	The Hong--Ou--Mandel (HOM) effect is often introduced through a single
	benchmark: coincidence suppression for \(\ket{1}\otimes\ket{1}\) at a
	balanced beam splitter. We present a classroom-oriented instructional module
	that broadens this treatment by comparing three output metrics---on/off
	coincidence probability \(P_c\), cross-correlation \(g^{(2)}_{12}\), and
	noise-reduction factor NRF---across four input families: Fock,
	superposition, coherent, and squeezed light. The module targets three
	instructional challenges in upper-division quantum optics: treating
	interference quality as a single observable, weakly connecting
	quantum--classical distinctions to output statistics, and limiting transfer
	from two-mode HOM reasoning to broader photonic benchmarking questions. A
	common beam-splitter sweep shows that different metrics probe different
	statistical properties of the same output state and can therefore support
	different source choices depending on the physical objective. The module
	combines a Jupyter/QuTiP simulator, guided activities, specification-style
	summary tables, and a grading rubric for use in a single upper-division or
	early graduate class meeting.
\end{abstract}
	
	\maketitle
	
	\section{Introduction}
	
	Quantum superposition, measurement, and interference are among the most conceptually challenging topics for students encountering quantum mechanics for the first time. Physics education research has shown that upper-division students often struggle to connect formal calculations to physical interpretation, especially in topics involving probability, measurement, and state discrimination.\cite{Singh2015-qu,Baily2010-yv,Bouchee2022-nf,Passante2015-ll} Related work on Quantum Interactive Learning Tutorials (QuILTs) has shown that guided, active-learning materials can help students connect quantum formalism with conceptual interpretation without sacrificing technical
	content.\cite{Singh2008-eg}
	In quantum optics, these challenges are amplified because students must reason not only about input states and measurements, but also about how different observables encode different aspects of the same optical transformation.\cite{justice2025framework}
	
	The Hong--Ou--Mandel (HOM) effect provides a natural setting for addressing these difficulties. As a canonical example of two-photon interference, HOM links amplitude-level reasoning to a directly measurable signature: suppression of coincidence events at a balanced beam splitter.\cite{Hong1987-fx} 
	
	At the same time, HOM is foundational for modern quantum photonics, where two-mode interference supports larger interferometric networks and modal encodings used in quantum communication, photonic information processing, and Gaussian Boson Sampling.\cite{O-Brien2009-ne,Hamilton2017-xo,Brecht2015-nf}
	
	Undergraduate laboratory implementations of the HOM interferometer naturally
	emphasize the canonical signature of coincidence suppression for
	$\ket{1}\otimes\ket{1}$ inputs at a balanced beam splitter.\cite{carvioto2012hong}
	In many instructional settings, however, HOM is therefore encountered primarily
	through this single observable. While important, that treatment leaves three
	recurring instructional challenges only weakly addressed:
	
	\begin{itemize}[leftmargin=*]
		\item \textbf{B1:} association of interference quality with a single observable;
		\item \textbf{B2:} weak connection between quantum--classical distinctions and the statistical meaning of specific output observables;
		\item \textbf{B3:} limited transfer from two-mode HOM reasoning to broader photonic benchmarking questions.
	\end{itemize}
	
	The present work addresses these challenges through a classroom-oriented
	course module rather than through a simulator alone. Its purpose is both practical and conceptual: to help instructors and students navigate the gap between HOM foundations and technology-facing photonic reasoning more easily.\cite{Bonilla-Licea2024-vx,Asfaw2022-le} We facilitate this process by providing a compact two-mode simulator, guided activities, specification-style summary tables, and rubric-based assessment. The core instructional claim is that the usefulness of a photonic state depends on the physical objective and therefore on the output metric used to judge performance.
	
This manuscript is written as a design-and-implementation paper rather than as a classroom-outcomes study. Its contribution is the coordinated packaging of a simulator, theory overview, activity sequence, assessment structure, and supporting materials into a form suitable for direct classroom use with limited additional development. The main text emphasizes the conceptual structure and classroom workflow, while the Supplementary Material and open-source notebook provide full implementation prompts, answer-key detail, mathematical results, and code-level transparency.
	
	\section{Prior work and design framework}
	\label{sec:background}
	
	The HOM effect has long served as a foundational example of two-particle
	quantum interference in both research and instructional settings.\cite{Hong1987-fx,Gerry2024-gc}
	Traditional treatments often emphasize operator-level derivations of
	coincidence suppression at a balanced beam splitter. While elegant, these
	derivations can be difficult for early learners when the algebra obscures the physical interpretation.

Educational simulations and laboratory activities have helped student learning
\cite{Mykhailova2020-rz,Deslauriers2011-lf,Krijtenburg-Lewerissa2017-aj,Muller2002-vg} and have made interference and measurement more accessible in instructional
settings. Examples include undergraduate photon and beam-splitter experiments and
more recent HOM implementations.\cite{holbrow2002photon,carvioto2012hong,DiBrita2023-wm}
	However, many tools either remain at the level of single-observable
	demonstrations or assume prior familiarity with advanced quantum optics
	software.\cite{McKagan2008-xu,DiBrita2023-wm,Bjurlin2025-fx}
	
	The present module occupies an intermediate position. It is more structured
	than a stand-alone simulator and more classroom-structured than a research notebook.
	Its design is organized around the instructional challenges identified in the
	Introduction and around backward-mapped learning goals rather than around
	software features alone. The intended contribution is therefore not only a
	simulator, but a classroom-oriented package that connects state choice, device
	operation, and metric-based technology reasoning. This packaging is intended to
	reduce instructor preparation time, a known barrier to instructional
	adoption.\cite{Folkers2026-po,Aga2024-gu}
	
	These design choices motivate five learning goals. After the module, students
	should be able to: (LG1) distinguish quantum from classical interference at a
	beam splitter using amplitude reasoning; (LG2) explain why coincidence
	probability, cross-correlation, and noise-reduction factor need not agree;
	(LG3) use multiple metrics to select input states for objective-dependent
	applications; (LG4) interpret engineering-style specification summaries as
	compact verification tools; and (LG5) connect two-mode HOM reasoning to larger
	photonic systems.
	
\begin{figure}[!h]
		\centering
		\includegraphics[width=0.8\linewidth]{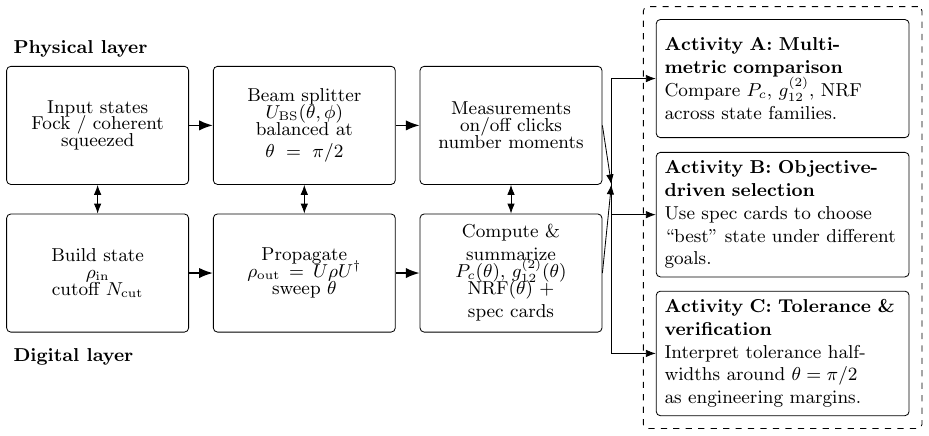}
		\caption{Graphical overview of the module. A common beam-splitter sweep is used to compare several input-state families under multiple observables. The same outputs are represented as plots for scientific interpretation and as specification-style summaries for engineering-style reasoning, with activities and assessment aligned to stated instructional challenges.}
		\label{fig:graphical}
	\end{figure}

	Figure~\ref{fig:graphical} provides a compact overview of the module design. It
	shows how a common beam-splitter sweep is used to compare multiple photonic
	input-state families under several output observables, while also emphasizing
	the module's multi-representation structure: scientific plots for conceptual
	interpretation, engineering-style specification summaries for decision-oriented
	reasoning, and aligned activities and assessment for classroom use. This
	overview is intended to orient the reader before the individual design elements
	are introduced in more detail below.
	
	Table~\ref{tab:per_to_design} summarizes how the challenges map to module elements and assessment evidence.
	
	\begin{table*}[t]
		\caption{Mapping from instructional challenge IDs (B1, B2, and B3) to design responses and assessment evidence. Activity labels A--C refer to the implementation summarized in Sec.~\ref{sec:implementation}.}
		\label{tab:per_to_design}
		\begin{tabular}{@{}l@{\hspace{0.012\textwidth}}l@{\hspace{0.025\textwidth}}l@{\hspace{0.025\textwidth}}l@{\hspace{0.025\textwidth}}l@{}}
			\hline \\[-0.1cm]
			\parbox[t]{0.08\textwidth}{\textbf{ID}} &
			\parbox[t]{0.20\textwidth}{\textbf{Instructional difficulty}} &
			\parbox[t]{0.20\textwidth}{\textbf{Design response}} &
			\parbox[t]{0.18\textwidth}{\textbf{Where implemented}} &
			\parbox[t]{0.22\textwidth}{\textbf{Learning goals and aligned evidence} \\[0.2cm]} \\
			\hline \\[-0.1cm]
			
			\parbox[t]{0.05\textwidth}{B1} &
			\parbox[t]{0.20\textwidth}{\justifying\noindent Association of interference quality with coincidence suppression alone.\cite{Gerry2024-gc,Borish2023-jz}} &
			\parbox[t]{0.20\textwidth}{\justifying\noindent Side-by-side comparison of $P_c$, $g^{(2)}_{12}$, and NRF across multiple state families.} &
			\parbox[t]{0.18\textwidth}{\justifying\noindent Activity A; Activity B; specification tables.} &
			\parbox[t]{0.22\textwidth}{\justifying\noindent \textbf{LG2, LG3:} metric-disagreement annotation and objective-driven state selection.} \\[4pt]
			
			\parbox[t]{0.05\textwidth}{B2} &
			\parbox[t]{0.20\textwidth}{\justifying\noindent Difficulty connecting quantum--classical distinctions to the statistical meaning of observables.\cite{Bouchee2022-nf,Singh2015-qu}} &
			\parbox[t]{0.20\textwidth}{\justifying\noindent Amplitude-based HOM reasoning contrasted with coherent-state baselines and correlation metrics.} &
			\parbox[t]{0.18\textwidth}{\justifying\noindent Theory; Activity A.} &
			\parbox[t]{0.22\textwidth}{\justifying\noindent \textbf{LG1, LG2:} amplitude explanation of HOM suppression and explicit Fock--coherent comparison.} \\[4pt]
			
			\parbox[t]{0.05\textwidth}{B3} &
			\parbox[t]{0.20\textwidth}{\justifying\noindent Weak transfer from two-mode HOM reasoning to larger photonic settings.\cite{O-Brien2009-ne,Hamilton2017-xo}} &
			\parbox[t]{0.20\textwidth}{\justifying\noindent Engineering-style specification summaries, tolerance prompts, and transfer questions.} &
			\parbox[t]{0.18\textwidth}{\justifying\noindent Activity B; Activity C; Discussion.} &
			\parbox[t]{0.22\textwidth}{\justifying\noindent \textbf{LG4, LG5:} acceptance-margin reasoning and extension to larger interferometric contexts. \\[-0.1cm]} \\
			\hline
		\end{tabular}
	\end{table*}
	
	\section{Theory overview}
	\label{sec:theory}
	
	The goal of this section is not to provide a complete treatment of HOM interferometry, but to introduce the minimal framework needed to interpret the simulator outputs and to support immediate use of the module. Readers who want a deeper formal account of derivations and exact identities are referred to the Supplementary Material Sec.~S1 and more specialized literature.\cite{Gerry2024-gc}
	
	\subsection{Photonic input states}
	The simulator compares four commonly encountered classes of two-mode photonic
	input states: Fock, superposition, coherent, and squeezed inputs. We use the
	standard shorthand $\ket{n,m}\equiv\ket{n}_a\ket{m}_b$ for two-mode number
	states.
	
	The Fock input $\ket{1}_a\ket{1}_b$ provides the canonical HOM benchmark.
	Because it has a definite photon number in each mode, it is the natural
	starting point for two-photon interference, coincidence suppression, and
	photon-counting logic. The product superposition state
	\begin{equation}
		\left(\frac{\ket{0}+\ket{1}}{\sqrt2}\right)_a
		\left(\frac{\ket{0}+\ket{1}}{\sqrt2}\right)_b
	\end{equation}
	provides an accessible nonclassical input with mixed vacuum and one-photon
	sectors. It is useful pedagogically because it remains easy to read in Dirac
	notation while already showing that two inputs can share one interference
	signature and differ under another.

	The coherent-state input serves as the classical-like optical reference. In
	practical terms, it models ideal laser light and provides the natural baseline
	for classical interferometry and alignment.

	The default squeezed-state case uses equal single-mode squeezed-vacuum inputs with fixed squeezing phase and
	user-selectable squeezing strength; this choice is intended as a fluctuation-suppressed continuous-variable reference rather than as a complete survey of squeezed-state interferometry. Formal
	definitions of coherent and squeezed states are standard and are not reproduced
	here; we refer the reader to Ref.~\cite{Gerry2024-gc} for details.
	
	Taken together, these four state families span the conceptual and technological
	range needed for the module: discrete single-photon inputs, simple nonclassical
	superpositions, classical-like optical baselines, and noise-reducing
	continuous-variable resources.
	
	\subsection{Beam splitter as a two-mode device primitive}
	
	The beam splitter is the basic two-mode photonic device primitive used throughout the module. It mixes two input modes into two output modes according to
	\begin{equation}
		\begin{pmatrix}
			\hat c\\
			\hat d
		\end{pmatrix}
		=
		\begin{pmatrix}
			t & r\\
			-r^* & t^*
		\end{pmatrix}
		\begin{pmatrix}
			\hat a\\
			\hat b
		\end{pmatrix},
		\qquad |t|^2+|r|^2=1,
		\label{eq:bs_transform_main}
	\end{equation}
	where $t$ and $r$ are complex transmission and reflection amplitudes. Their squared magnitudes give the transmitted and reflected fractions. In the simulator we use the following parameterization (Ref. \cite{Gerry2024-gc})
	\begin{equation}
		t=\cos\!\left(\frac{\theta}{2}\right),\qquad
		r=e^{i\phi}\sin\!\left(\frac{\theta}{2}\right),
	\end{equation}
	where $\theta$ controls the mixing ratio and $\phi$ the relative phase. The balanced $50{:}50$ operating point is
	\begin{equation}
		\theta=\frac{\pi}{2},
	\end{equation}
	which is the main HOM setting used throughout the module.
	
	This formulation is intentionally minimal. For present purposes, the key point is that the beam splitter does not merely route light: it coherently mixes the two modes. As a result, the output statistics depend on both the input state and the device setting $(\theta,\phi)$. That dependence is the basis of quantum interference in the present two-mode setting.
	Table~\ref{tab:default_params} lists the default simulator settings used for
	the examples and activity prompts. These values define the reference module
	configuration; instructors can vary them in the notebook to construct
	extensions or homework variants.
	\begin{table}[t]
		\caption{Default simulator settings used in the module examples.}
		\label{tab:default_params}
		\begin{tabular}{ll}
			\hline
			\textbf{Quantity} & \textbf{Default value} \\
			\hline
			Beam-splitter phase & \(\phi=\pi/2\) \\
			Beam-splitter sweep & \(\theta\in[0,\pi]\), balanced at \(\theta=\pi/2\) \\
			Coherent input & \(\ket{\alpha}_a\ket{\alpha}_b\), \(|\alpha|=1\), \(\arg\alpha=0\) \\
			Squeezed input & equal SMSV inputs, \(r=6\,\mathrm{dB}\) \\
			Squeezing phases & \(\phi_{\mathrm{sq},1}=0\), \(\phi_{\mathrm{sq},2}=\pi/2\) \\
			Fock cutoff & \(N_{\mathrm{cut}}=12\) \\
			Default sweep grid & 61 \(\theta\)-points \\
			\hline
		\end{tabular}
	\end{table}
	
\subsection{From input states to output-state measurements}

The beam splitter acts on a two-mode input state and produces a corresponding two-mode output state. One may think of this process in state language: an input state $\ket{\psi}_{ab}$ defined on modes $a$ and $b$ is transformed into an output state $\ket{\psi_{\mathrm{out}}(\theta)}_{cd}$ defined on modes $c$ and $d$. In the idealized settings emphasized in this module, this state-based picture is sufficient for understanding how different photonic inputs lead to different output statistics. To connect this state description to measurement, we use the photon-number operators
\begin{equation}
	\hat n_c=\hat c^\dagger \hat c,
	\qquad
	\hat n_d=\hat d^\dagger \hat d,
\end{equation}
which count photons in the two output modes. Here $\hat c^\dagger,\hat c$ and $\hat d^\dagger,\hat d$ are the usual creation and annihilation operators for the output modes.\cite{Gerry2024-gc} These photon-number operators provide the link between the output state and the measurable quantities reported by the simulator. In particular, the module focuses on observables built from photon-number events, photon-number correlations, and photon-number fluctuations at the outputs. For a pedagogical introduction, these can be understood as three
different ways of assessing the beam splitter action on the input light: how often both outputs click together, how strongly the outputs are correlated, and how noisy the output-number difference is.

For the classroom use of the module, students do not need to manipulate density
matrices explicitly. The essential idea is that the beam splitter produces an
output state, and the simulator then asks three measurement questions about
that state: what is the probability that both threshold detectors click, what is
the normalized photon-number correlation between the two ports, and how large
are the fluctuations in the output-number difference.

Internally, the simulator represents the output by a two-mode density matrix $\rho_{cd}(\theta)$. This is a convenient computational representation because the same notation can evaluate probabilities, expectation values, and fluctuations, and it can later be extended to mixed states, loss, or partial distinguishability. For the pure-state examples emphasized here, however, $\rho_{cd}(\theta)$ may simply be read as the state information needed to compute the output measurement statistics.
	
\subsection{Three operational output metrics}

The module uses three complementary output metrics, each corresponding to a
different measurement question about the two beam-splitter outputs. The first
metric is the on/off coincidence probability,
\begin{equation}
	\begin{aligned}
		P_c(\theta)
		&=
		1-P_{0,c}(\theta)-P_{0,d}(\theta)+P_{0,cd}(\theta),
		\\
		P_{0,c}(\theta)
		&=
		\mathrm{Tr}\!\left[
		\rho_{cd}(\theta)
		\left(\ket{0}\!\bra{0}_c\otimes I_d\right)
		\right],
		\\
		P_{0,d}(\theta)
		&=
		\mathrm{Tr}\!\left[
		\rho_{cd}(\theta)
		\left(I_c\otimes \ket{0}\!\bra{0}_d\right)
		\right],
		\\
		P_{0,cd}(\theta)
		&=
		\mathrm{Tr}\!\left[
		\rho_{cd}(\theta)\ket{0,0}\!\bra{0,0}
		\right].
	\end{aligned}
	\label{eq:pc_main}
\end{equation}
This is the probability that both threshold detectors register at least one
photon. For the \(\ket{1,1}\) input, this on/off coincidence probability
coincides with the usual one-photon-in-each-output HOM coincidence event.
For coherent and squeezed inputs, it includes multiphoton threshold-detector
coincidences and should therefore be interpreted as an on/off detection
metric rather than as a projection only onto \(\ket{1,1}\).
The symbol \(\rho_{cd}(\theta)\) denotes the two-mode output density matrix
after the beam splitter.
	
	The second metric is the normalized cross-correlation,
	\begin{equation}
		g^{(2)}_{12}(\theta)=
		\frac{\langle \hat n_c\hat n_d\rangle}
		{\langle \hat n_c\rangle\langle \hat n_d\rangle},
		\label{eq:g2_main}
	\end{equation}
	where $\hat n_c$ and $\hat n_d$ are the photon-number operators for the two
	output modes. This metric measures how strongly the two output photon numbers
	are correlated relative to the product of their mean intensities. It also
	provides a useful classical reference: for equal coherent inputs in the present
	setting, $g^{(2)}_{12}=1$.
	
	The third metric is the noise-reduction factor,
	\begin{equation}
		\mathrm{NRF}(\theta)=
		\frac{\mathrm{Var}(\hat n_c-\hat n_d)}
		{\langle \hat n_c+\hat n_d\rangle}.
		\label{eq:nrf_main}
	\end{equation}
	NRF measures fluctuations in the output-number difference relative to the total
	mean output photon number, giving a shot-noise-style benchmark for
	number-difference noise.
	
	These three metrics correspond to three different technology-facing questions.
	Coincidence probability is natural for coincidence-based logic and Bell-type
	measurements.\cite{Hong1987-fx,O-Brien2009-ne} Cross-correlation is useful for
	source characterization and for distinguishing classical-like from nonclassical
	output statistics.\cite{Gerry2024-gc} Noise-reduction factor is natural for
	interferometric sensing and sub-shot-noise applications.\cite{Hamilton2017-xo}
	Their inclusion side by side is therefore central to the module design: they
	probe different statistical properties of the same transformed output state and
	need not rank sources in the same way.
	
	The central comparison made visible by the simulator is that no input family is
	universally optimal. The Fock input $\ket{1}\otimes\ket{1}$ and the product
	superposition input $\tfrac{1}{\sqrt2}(\ket0+\ket1)^{\otimes 2}$ both suppress
	coincidences at the balanced beam splitter, but their coincidence curves differ
	in scale. Their $g^{(2)}_{12}$ curves are identical in the idealized model, so
	that metric alone cannot distinguish them. Their NRF curves, however, cross
	away from the balanced point, showing that the preferred input can depend on
	the operating region. 
	For the equal-amplitude coherent input \(\ket{\alpha}_a\ket{\alpha}_b\)
	under the default quadrature beam-splitter phase \(\phi=\pi/2\), the output
	coherent amplitudes retain equal magnitudes throughout the sweep; consequently
	\(P_c\), \(g^{(2)}_{12}\), and NRF are flat classical-like references in the
	simulator. Squeezed inputs can reduce number-difference noise without
	optimizing the coincidence metric. State selection for a photonic application must therefore
	be driven by both the relevant output metric and the operating point at which
	the device will be used.
	
\section{Simulator and access layers}
\label{sec:simulator}
The simulator is an interactive optical workbench for exploring two-mode
interference in a common beam splitter parameterization. Each run follows a
simple workflow: select one or more input-state families, sweep the beam
splitter mixing angle $\theta$ through the balanced point
$\theta=\pi/2$, and compare the resulting $P_c(\theta)$,
$g^{(2)}_{12}(\theta)$, and NRF$(\theta)$ curves.

The interface is designed to support three levels of engagement. At the visual
level, full metric sweeps allow students to compare curve shapes, identify
extrema, and observe that different inputs can behave differently under the
same optical transformation.\cite{McKagan2008-xu} At the operational level,
specification-style summaries report minima, maxima, operating-point values,
and tolerance half-widths, allowing students to make engineering-style
decisions. At the computational level, the open-source Jupyter/QuTiP notebook
allows interested users to inspect state construction, beam splitter
propagation, and metric calculations directly.

For a low-preparation classroom implementation, an instructor can begin with
one or two preset state families, sweep $\theta$ through the balanced point, and use the metric
plots and specification summaries to guide discussion. The simulator is
implemented in Python using QuTiP in a truncated two-mode Fock basis. Input
states are propagated through the beam splitter transformation and evaluated
under the three output metrics defined above. The same implementation can be
extended naturally to mixed states, loss, or partial distinguishability. For
simulator setup and first-use guidance, see Supplementary Material Sec.~S2.

\begin{figure}[t]
	\centering
	\includegraphics[width=0.82\linewidth]{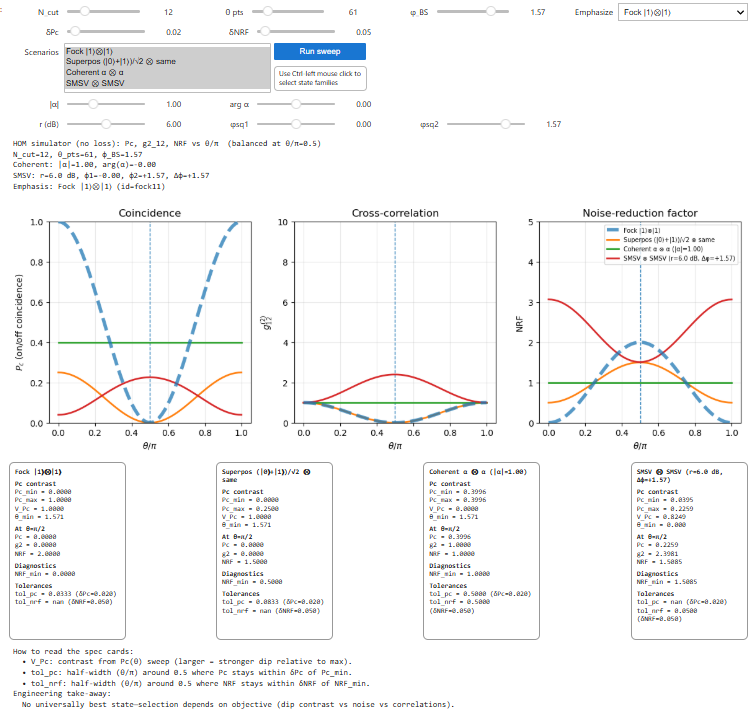}
	\caption{Representative layout of the simulator. The interface combines
		state selection, beam splitter sweeps, multi-metric plots, and
		specification-style summaries so that the same physical calculation can be
		interpreted visually, operationally, and computationally.}
	\label{fig:simulator}
\end{figure}

\section{Module implementation}
\label{sec:implementation}

The module is designed for a single 50-minute class meeting in an upper-division undergraduate or early graduate quantum optics course (see also Supplementary Material Secs.~S3 and~S4). It uses a predict--observe--explain structure: students first record a written prediction, then interact with the simulator, and finally write a brief explanation reconciling the two. This structure is consistent with evidence that hypothesis formation before observation supports conceptual engagement.\cite{Crouch2001-hh,McKagan2008-xu}

A typical implementation uses a brief instructor-led discussion of the Fock and coherent cases, a paired metric-selection activity using specification tables, and a short transfer prompt connecting the two-mode module to a larger photonic network. In practice, the module is intentionally adjustable: an instructor can use all three activity blocks in a full 50-minute format or shorten the implementation by trimming one activity while retaining the same underlying trade-off structure.

\paragraph*{Compact classroom workflow.}
The activity sequence is summarized below; full prompts, instructor notes, concept checks, and answer-key details are provided in Supplementary Material Secs.~S5--S7.

\begin{description}[leftmargin=*,labelsep=0.5em]
	
	\item[Activity A: Multi-metric HOM comparison (B1, B2)]
	Students compare Fock, superposition, coherent, and squeezed inputs under
	\(P_c\), \(g^{(2)}_{12}\), and NRF. They record one example of metric
	agreement and one example of metric disagreement. The main learning point is
	that interference quality is observable-dependent and that \(P_c\) alone is
	not a complete benchmark. Suggested time: 10--15 min.
	
	\item[Activity B: Metric-driven state selection (B1, B3)]
	Students use specification-style summaries to choose an input family for
	different objectives. They record a state choice, a metric-based
	justification, and one trade-off. The main learning point is that the
	preferred source depends on the metric and operating point. Suggested time:
	10 min.
	
	\item[Activity C: Robustness and transfer (B3)]
	Students interpret tolerance half-widths and discuss embedding the same
	two-mode primitive in a larger network. They record an acceptance-window
	statement and one network-level extension. The main learning point is that
	two-mode HOM reasoning can become a local verification tool for larger
	photonic systems. Suggested time: 5--8 min.
	
\end{description}
	
	\section{Assessment}
	\label{sec:assessment}
	
	Assessment is built at two levels: brief concept checks and matched implementation prompts (Supplementary Material Sec.~S7). In the present manuscript we emphasize the rubric (Tab.~\ref{tab:rubric} and Supplementary Material Sec.~S8) because it is the most compact representation of what the module measures and how it aligns with the stated learning goals (LG1 to LG5). 
	
	\begin{table*}[h]
		\caption{Grading rubric for the module (10 points total). LG4 is required for the full implementation; it may be omitted in compressed modes without penalty to the remaining criteria.}
		\label{tab:rubric}
		\footnotesize
		\renewcommand{\arraystretch}{1.08}
		\setlength{\tabcolsep}{4pt}
		\begin{tabular}{ll}
			\hline
			\textbf{Criterion} & \textbf{Performance levels} \\
			\hline
			
			LG1 (0--2 pts) &
			\parbox[t]{0.78\textwidth}{Quantum--classical distinction:\ correct amplitude-cancellation explanation contrasted with coherent baseline (2);\ amplitude language present but comparison incomplete or incorrect (1);\ incorrect or missing (0).} \\[4pt]
			
			LG2 (0--3 pts) &
			\parbox[t]{0.78\textwidth}{Metric dependence:\ correctly identifies that $P_c$, $g^{(2)}_{12}$, and NRF probe different statistical properties and gives a specific example of disagreement with justification (3);\ identifies disagreement without statistical explanation or with an incomplete example (2);\ asserts disagreement without supporting reasoning (1);\ incorrect (0).} \\[4pt]
			
			LG3 (0--2 pts) &
			\parbox[t]{0.78\textwidth}{Objective-driven selection:\ selects a state family with evidence matched to the stated objective and explicitly names the trade-off with at least one other metric (2);\ correct selection without trade-off reasoning (1);\ incorrect or no justification (0).} \\[4pt]
			
			LG4 (0--2 pts) &
			\parbox[t]{0.78\textwidth}{Verification reasoning:\ uses tolerance half-width and threshold to formulate a clear acceptance or rejection statement with correct numerical comparison (2);\ uses specification-table entries but comparison is incomplete (1);\ incorrect or missing (0).} \\[4pt]
			
			LG5 (0--1 pt) &
			\parbox[t]{0.78\textwidth}{Transfer:\ gives a credible argument connecting two-mode HOM reasoning to a larger photonic or interferometric context, naming at least one specific extension required (1);\ incomplete or no transfer argument (0).} \\
			\hline
		\end{tabular}
	\end{table*}
	
	This assessment structure supports multiple audiences. A student who uses the module mainly at the visual level can still engage meaningfully with LG1 and LG2. A student or instructor interested in technology-facing reasoning can engage more directly with LG3 and LG4 through the specification tables. A more advanced reader can use the computational layer and appendix to connect these judgments back to state representation and exact identities.
	
	\section{Scope and contribution}
	\label{sec:discussion}
	
	The module deliberately excludes multi-mode entanglement, detector dark counts, timing jitter, mode mismatch, and full loss channels. These are not oversights but intentional omissions: each would require formal machinery beyond a single-class two-mode module. They also provide a natural path for future extensions. In particular, partial distinguishability, explicit loss, and network-level tolerance propagation are especially natural follow-ons because they preserve the same basic logic of state preparation, device action, and metric-based evaluation.
	
	Within that scope, the paper contributes a classroom-oriented example of challenge-targeted instructional design in upper-division quantum optics. Its value lies in the coordinated packaging of a simulator, state comparisons, engineering-style outputs, aligned assessment, and curated supplementary materials into a form that supports low-barrier entry without removing a path to deeper study. The module is designed so that a reader can begin at the level of physical interpretation and technology selection, while more advanced users can continue into code inspection, density-matrix reasoning, and analytic derivations.
	
	The central design lesson is that state selection for a photonic application must be driven by the metric corresponding to the technological objective. Making that trade-off structure visible in a lightweight, notebook-native environment provides a practical bridge from HOM foundations to broader questions of benchmarking and design in quantum photonics.
	
	\section{Conclusion}
	\label{sec:conclusion}
	
	We have presented a classroom-oriented HOM module for upper-division quantum optics built around three instructional challenges, rather than around simulator features alone. The module uses a common two-mode beam splitter sweep to compare four photonic input families under three output metrics, making visible that coincidence probability, cross-correlation, and noise-reduction factor answer different physical and technological questions. The resulting trade-offs are expressed both visually and through engineering-style specification summaries, allowing the same simulator to support quick entry, classroom deployment, engineering-style reasoning, and deeper technical inspection. In this way, the module provides a compact bridge between HOM foundations and technology-facing photonic reasoning.

	\begin{acknowledgements}
		M.C.A.\ and B.K.\ gratefully acknowledge support from the National Science Foundation through the QCAP-Pilot and QCAP-Design efforts under NSF Award Nos.~OSI-2410813 and OSI-2531569.
	\end{acknowledgements}
	
	\section*{Conflict of interest}
	The authors have no conflicts to disclose.
	
	\section*{Author contributions}
	M.C.A. contributed to simulator testing, verification of instructional outputs,
	and manuscript editing. B.K. conceived the project, developed the pedagogical
	framing, carried out the primary theoretical and manuscript development, and
	supervised the work.
	
	\section*{Data availability}
	The HOM simulator notebook and supporting materials are available at the GitHub repository: \url{https://github.com/boriskiefer/sim-HOM}.
	
	\bibliographystyle{unsrt}
	\bibliography{sim_HOM}
	
\clearpage
\section*{Supplementary Material}

This Supplementary Material is organized to support both rapid instructional adoption and deeper technical inspection. It contains extended mathematical details, simulator setup guidance, an instructor implementation guide, implementation variants, full activity prompts, instructor notes, concept checks, pre/post prompts, and assessment guidance.

\subsection*{S1. Extended mathematical details}
\label{math_details}

This section collects exact formulas and supplementary explanations that support the qualitative comparisons made in the main paper.

\paragraph*{Key metric identities used in the module.}
The following identities are the analytic basis of the main instructional comparisons in Activities A and B. For the Fock input $\ket{1,1}$,
\begin{equation}
	P_c^{\mathrm{Fock}}(\theta)=\cos^2\theta,
	\qquad
	g_{12}^{(2),\mathrm{Fock}}(\theta)=\cos^2\theta,
	\qquad
	\mathrm{NRF}_{\mathrm{Fock}}(\theta)=2\sin^2\theta.
\end{equation}
For the superposition input $\ket{+}_a\ket{+}_b$ with $\ket{+}=(\ket0+\ket1)/\sqrt2$,
\begin{equation}
	P_c^{\mathrm{sup}}(\theta)=\frac14\cos^2\theta,
	\qquad
	g_{12}^{(2),\mathrm{sup}}(\theta)=\cos^2\theta,
	\qquad
	\mathrm{NRF}_{\mathrm{sup}}(\theta)=\frac12+\sin^2\theta.
\end{equation}
Thus
\begin{equation}
	g_{12}^{(2),\mathrm{sup}}(\theta)=g_{12}^{(2),\mathrm{Fock}}(\theta)
\end{equation}
for all $\theta$, while the NRF curves cross at
\begin{equation}
	\theta=\frac{\pi}{4}.
\end{equation}
For equal-amplitude coherent inputs \(\ket{\alpha}_a\ket{\alpha}_b\) under the default quadrature beam-splitter phase \(\phi=\pi/2\), the output coherent amplitudes retain equal magnitudes throughout the sweep. Therefore the three simulator metrics are flat in this reference configuration:
\begin{equation}
	P_c^{\mathrm{coh}}=\mathrm{const},\qquad
	g_{12}^{(2),\mathrm{coh}}=1,\qquad
	\mathrm{NRF}_{\mathrm{coh}}=1.
\end{equation}

Here \(P_c\) denotes the on/off coincidence probability defined in
Eq.~\eqref{eq:pc_main}. For non-number-state inputs this differs from the exact projection probability onto \(\ket{1,1}\), because threshold
coincidences include multiphoton events.
These identities explain the main qualitative comparisons emphasized in the module: scaled coincidence dips for Fock and superposition inputs, structural degeneracy in $g^{(2)}_{12}$, an operating-point-dependent NRF preference, and coherent-state flatness as a robustness baseline. 

\subsection*{S2. Simulator setup and first-use guidance}

The simulator is implemented as an open-source Jupyter/QuTiP notebook. A recommended first use is:
\begin{enumerate}[leftmargin=*]
	\item select the Fock and coherent input families,
	\item sweep $\theta$ through the balanced point,
	\item compare $P_c(\theta)$, $g^{(2)}_{12}(\theta)$, and NRF$(\theta)$,
	\item inspect the specification-style summaries,
	\item then add the superposition and squeezed inputs.
\end{enumerate}

For first-time users, the most important interpretation step is to compare how the \emph{same} beam splitter sweep supports \emph{different} judgments depending on which output metric is taken as primary. This is the central instructional idea of the module.

\subsection*{S3. Instructor implementation guide}

This module is intended for upper-division undergraduate or early graduate instruction in quantum optics or related courses. It is designed for a single 50-minute class meeting, but may also be adapted for shorter in-class use or homework-based follow-up.

\paragraph*{Recommended prerequisites.}
Students should be familiar with basic Dirac notation, optical beam splitters at a qualitative level, and the distinction between quantum states and observables. Prior exposure to second-quantized notation is helpful but not strictly required for the main classroom implementation.

\paragraph*{Learning goals.}
The module supports the five learning goals stated in the main manuscript:
\begin{itemize}[leftmargin=*]
	\item LG1: distinguish quantum from classical interference at a beam splitter using amplitude reasoning;
	\item LG2: explain why coincidence probability, cross-correlation, and noise-reduction factor need not agree;
	\item LG3: use multiple metrics to select input states for objective-dependent applications;
	\item LG4: interpret engineering-style specification summaries as compact verification tools;
	\item LG5: connect two-mode HOM reasoning to larger photonic systems.
\end{itemize}

\paragraph*{Recommended classroom flow (50 minutes).}
A typical deployment is:
\begin{enumerate}[leftmargin=*]
	\item warmup prediction or pre-prompt (5 minutes);
	\item short instructor framing of Fock and coherent inputs at the balanced beam splitter (10 minutes);
	\item Activity A, multi-metric HOM comparison (15 minutes);
	\item Activity B, metric-driven state selection with specification tables (10 minutes);
	\item Activity C, robustness and transfer (5 minutes);
	\item wrap-up and concept check (5 minutes).
\end{enumerate}

\paragraph*{Minimum viable implementation.}
If time is limited, the module can be reduced to:
\begin{itemize}[leftmargin=*]
	\item one brief prediction prompt,
	\item one simulator sweep comparing Fock and coherent inputs,
	\item one specification-table decision task,
	\item one short transfer question.
\end{itemize}
This compressed implementation still supports LG1--LG3 and introduces the central instructional claim that metric choice changes which source appears preferable.

\subsection*{S4. Implementation variants}

\paragraph*{Full 50-minute version.}
Use all three activities, one pre-prompt, and one concept check.

\paragraph*{Compressed 30-minute version.}
Use PP1, a short Fock/coherent comparison, one metric-selection task from Activity B, and one transfer question from Activity C.

\paragraph*{Homework or recitation variant.}
Assign the simulator sweep outside class, then use class time for Activity B and the concept checks.

\paragraph*{Advanced follow-up version.}
Use the appendix and simulator notebook to extend the discussion toward distinguishability, loss, or larger interferometric architectures.

\subsection*{S5. Full activity prompts}

\paragraph*{Activity A: multi-metric HOM comparison (B1, B2).}
\textbf{Purpose.} Compare the four input families under all three output metrics and identify where different metrics agree or disagree.

\textbf{Student prompt.}
Using the simulator, compare the Fock, superposition, coherent, and squeezed input families under the three output metrics $P_c(\theta)$, $g^{(2)}_{12}(\theta)$, and NRF$(\theta)$. At the balanced beam splitter, identify which state families appear favorable under each metric. Then examine the full sweep and note at least one example where two states agree under one metric but differ under another. Summarize your observations in three to four sentences.

\textbf{Expected student deliverable.}
A short comparison identifying at least one example of metric disagreement and one example of a classical-like baseline.

\textbf{Suggested timing.}
10--15 minutes.

\paragraph*{Activity B: metric-driven state selection (B1, B3).}
\textbf{Purpose.} Use engineering-style summaries to select the best input family for different objectives.

\textbf{Student prompt.}
Use the specification-style tables generated by the simulator to decide which input family you would choose for each of the following goals:
\begin{enumerate}[leftmargin=*]
	\item minimize coincidence probability at $\theta=\pi/2$,
	\item minimize NRF near the balanced operating point,
	\item maximize tolerance half-width for a chosen threshold.
\end{enumerate}
For each choice, justify your selection using at least one metric and one trade-off relative to another state family.

\textbf{Expected student deliverable.}
A short objective-by-objective state selection with explicit metric-based justification.

\textbf{Suggested timing.}
10 minutes.

\paragraph*{Activity C: robustness and transfer (B3).}
\textbf{Purpose.} Extend two-mode reasoning toward device screening and larger photonic systems.

\textbf{Student prompt.}
The simulator reports tolerance half-widths around a chosen operating point. Explain how a tolerance half-width can function as an acceptance criterion in a photonic device specification. Then discuss how this reasoning would need to be extended if the same two-mode beam splitter were embedded inside a larger interferometric network.

\textbf{Expected student deliverable.}
A brief explanation of tolerance-based acceptance reasoning plus one concrete transfer statement to a larger system.

\textbf{Suggested timing.}
5--8 minutes.

\subsection*{S6. Instructor notes and expected responses}

\paragraph*{Activity A notes.}
Students often begin by treating $P_c$ as the only meaningful HOM benchmark. A productive intervention is to ask them whether a state that minimizes $P_c$ must also minimize fluctuations or maximize robustness. Good responses identify that the same beam splitter transformation can look favorable or unfavorable depending on the chosen metric.

\paragraph*{Activity B notes.}
Students may choose the ``best'' state without explicitly naming the objective. Encourage them to complete the sentence: \emph{This state is best if the device goal is \ldots\ because the relevant metric is \ldots} Full-credit responses should mention at least one trade-off.

\paragraph*{Activity C notes.}
Students often understand tolerance half-width qualitatively but do not immediately connect it to verification or acceptance language. Prompt them to rephrase the half-width as an allowed operating window around the design point. For transfer, strong answers note that larger interferometric networks require propagation of local tolerances through multiple optical elements.

\paragraph*{Concept-check notes.}
\begin{itemize}[leftmargin=*]
	\item CC1: Full-credit responses should use probability-amplitude cancellation for indistinguishable two-photon paths and should not assign the same mechanism to coherent light.
	\item CC2(a): Full-credit responses should explicitly state that equal $g^{(2)}_{12}$ does not imply physical equivalence, because $P_c$ and NRF differ.
	\item CC2(b): Good answers typically identify sensing or sub-shot-noise measurement as favorable for squeezed inputs, and coincidence-based logic or Bell-type applications as unfavorable if coincidence suppression is the key objective.
	\item CC3: Full-credit responses should compare the tolerance half-width with the fabrication spread and then identify at least one additional system-level requirement, such as cumulative tolerance propagation, network architecture, or multi-element calibration.
\end{itemize}

\subsection*{S7. Concept checks and pre/post prompts}

\paragraph*{Concept checks.}

\textbf{CC1 (LG1, LG2).} At $\theta=\pi/2$, the simulator reports $P_c(\pi/2)=0$ for the Fock input $\ket{1}\otimes\ket{1}$. Explain in two sentences why this suppression occurs using probability-amplitude language, and state whether you would expect the same suppression for a coherent input $\ket{\alpha}\otimes\ket{\alpha}$. Justify your answer.

\medskip
\textbf{CC2 (LG2, LG3).} The simulator reports the same $g^{(2)}_{12}(\theta)$ for the Fock input $\ket{1}\otimes\ket{1}$ and the superposition input $\frac{1}{\sqrt{2}}(\ket{0}+\ket{1})^{\otimes 2}$ at every beam-splitter angle.
\begin{enumerate}[label=(\alph*),leftmargin=*]
	\item Does this mean the two states are physically equivalent? Justify your answer by comparing $P_c(\theta)$ and NRF$(\theta)$ across the sweep.
	\item For a squeezed input, the simulator additionally reports NRF$(\pi/2) < 1$ but $g^{(2)}_{12}(\pi/2) > 1$. Identify a technological application for which the squeezed input would be preferred over the Fock input, and one for which it would not.
\end{enumerate}

\medskip
\textbf{CC3 (LG4, LG5).} The specification table reports a tolerance half-width of $\Delta\theta = 0.15$ rad for a stated NRF threshold. A fabrication process produces beam splitters with $\theta = \pi/2 \pm 0.20$ rad. Does this source meet the specification? State what additional information you would need to make this judgment for a four-port photonic network.

\paragraph*{Pre/post prompts.}

\textbf{PP1 (LG1).} Before running the simulator: predict whether you expect the coincidence probability $P_c(\pi/2)$ to be zero, nonzero, or undefined for a coherent input $\ket{\alpha}\otimes\ket{\alpha}$ at a balanced beam splitter. Justify your prediction in two sentences.

\medskip
\textbf{PP2 (LG2, LG3).} For an instructor-selected state family, run the three-metric sweep, record the values at $\theta=\pi/2$, and identify which metric, if any, is best satisfied by this state family.

\medskip
\textbf{PP3 (LG4, LG5).} Explain in three to four sentences how the tolerance half-width reported by the simulator could serve as an acceptance criterion in a quantum photonic device specification, and identify one way in which this reasoning would need to be extended for a larger interferometric system.

\subsection*{S8. Assessment guidance}

The rubric in the main text is intended for fast classroom grading. The following clarifications may help instructors apply it consistently.

\paragraph*{LG1.}
A full-credit response should explicitly distinguish quantum interference from classical-like averaging and should refer to amplitude cancellation or indistinguishable paths.

\paragraph*{LG2.}
A full-credit response should name at least two different metrics and explain that they probe different statistical properties, rather than merely stating that the curves look different.

\paragraph*{LG3.}
A full-credit response must match the selected state to the stated objective and mention at least one trade-off with another metric or state family.

\paragraph*{LG4.}
A full-credit response should use the specification table numerically or operationally rather than only qualitatively.

\paragraph*{LG5.}
A full-credit response should go beyond ``larger systems are more complicated'' and identify one concrete extension, such as propagation of local tolerances, cumulative loss, or architecture-dependent sensitivity.
	
\end{document}